\documentclass{ws-p10x7}

\begin{document}

\title{Constraints on the Proton's Gluon Density from Lepton-Pair Production}

\author{Edmond L. Berger}

\address{High Energy Physics Division, Argonne National Laboratory, 
Argonne, IL 60439,
USA\\E-mail: berger@anl.gov}

\author{M. Klasen}

\address{II.~Institut f\"ur Theoretische Physik, Universit\"at Hamburg, 
             D-22761 Hamburg, Germany\\E-mail: michael.klasen@desy.de}  

\twocolumn[\maketitle\abstract{Massive lepton-pair production, the Drell-Yan 
process, should be a good source of independent constraints on the gluon 
density, free from the experimental and theoretical complications of photon 
isolation that beset studies of prompt photon production.  We provide 
predictions for the spin-averaged and spin-dependent differential cross 
sections as a function of transverse momentum $Q_T$.   
}]
  \vspace*{-10.3cm}
  \noindent hep-ph/0009257 \\
  ANL-HEP-CP-00-098 \\
  DESY 00-135
  \vspace*{8.0cm}
\section{Introduction}

Massive lepton-pair production, $h_1 + h_2 \rightarrow \gamma^* + X;
\gamma^* \rightarrow l \bar{l}$, the Drell-Yan process,
and prompt real photon production, $h_1 + h_2 \rightarrow \gamma + X$, are two 
of the most valuable probes of short-distance behavior in hadron reactions.  
They supply critical information on parton momentum densities and 
opportunities for tests of perturbative quantum chromodynamics (QCD).
Spin-averaged parton momentum densities may be extracted from spin-averaged 
nucleon-nucleon reactions, and spin-dependent parton momentum densities from 
spin-dependent nucleon-nucleon reactions.  

The Drell-Yan process has tended to be thought of primarily as a source of 
information on quark densities.  Indeed, the mass and longitudinal momentum 
(or rapidity) dependences of the cross section (integrated over the transverse
momentum $Q_T$ of the pair) provide essential constraints on the 
{\it antiquark} momentum density, complementary to deep-inelastic lepton 
scattering from which one gains information of the sum of the quark and 
antiquark densities.  Prompt real photon production, on the other hand, is a 
source of essential information on the {\it gluon} momentum density.  At 
lowest order in perturbation theory, the reaction is dominated at large values 
of the transverse momentum $p_T$ of the produced photon by the QCD ``Compton" 
subprocess, $q + g \rightarrow \gamma + q$.  This dominance is preserved at 
higher orders, indicating that the experimental inclusive cross section 
differential in $p_T$ may be used to determine the density of gluons in the 
initial hadrons.

In this contribution, we summarize recent work\cite{BGKDY,BGKDYX}, in 
which we demonstrate that the QCD Compton subprocess, 
$q + g \rightarrow \gamma^* + q$
also dominates the Drell-Yan cross section in polarized and unpolarized 
proton-proton reactions for values of the transverse 
momentum $Q_T$ of the pair that are larger than roughly half of the pair 
mass $Q$, $Q_T > Q/2$.  Dominance of the $qg$ contribution in the massive 
lepton-pair case is as strong if not stronger than it is in the prompt photon 
case.  Massive lepton-pair differential cross sections are therefore an 
additional useful source of constraints on the the spin-averaged 
and spin-dependent {\it gluon densities}.  Although the cross section 
is smaller than the prompt photon cross section, massive lepton pair production 
is cleaner theoretically since long-range fragmentation contributions are absent 
as are the experimental and theoretical complications associated with isolation 
of the real photon.  As long $Q_T$ is large, the perturbative requirement of 
small $\alpha_s(Q_T)$ can be satisfied without a large value of $Q$.  We therefore 
explore and advocate the potential advantages of studies of 
$d^2\sigma/dQ dQ_T$ as a function of $Q_T$ for modest values of $Q$,  
$Q \sim 2$GeV, below the range of the traditional Drell-Yan region. 

\section{Unpolarized Cross Sections}

For $p \bar{p} \rightarrow \gamma^* + X$ at
\begin{figure}
\epsfxsize200pt
\figurebox{120pt}{160pt}{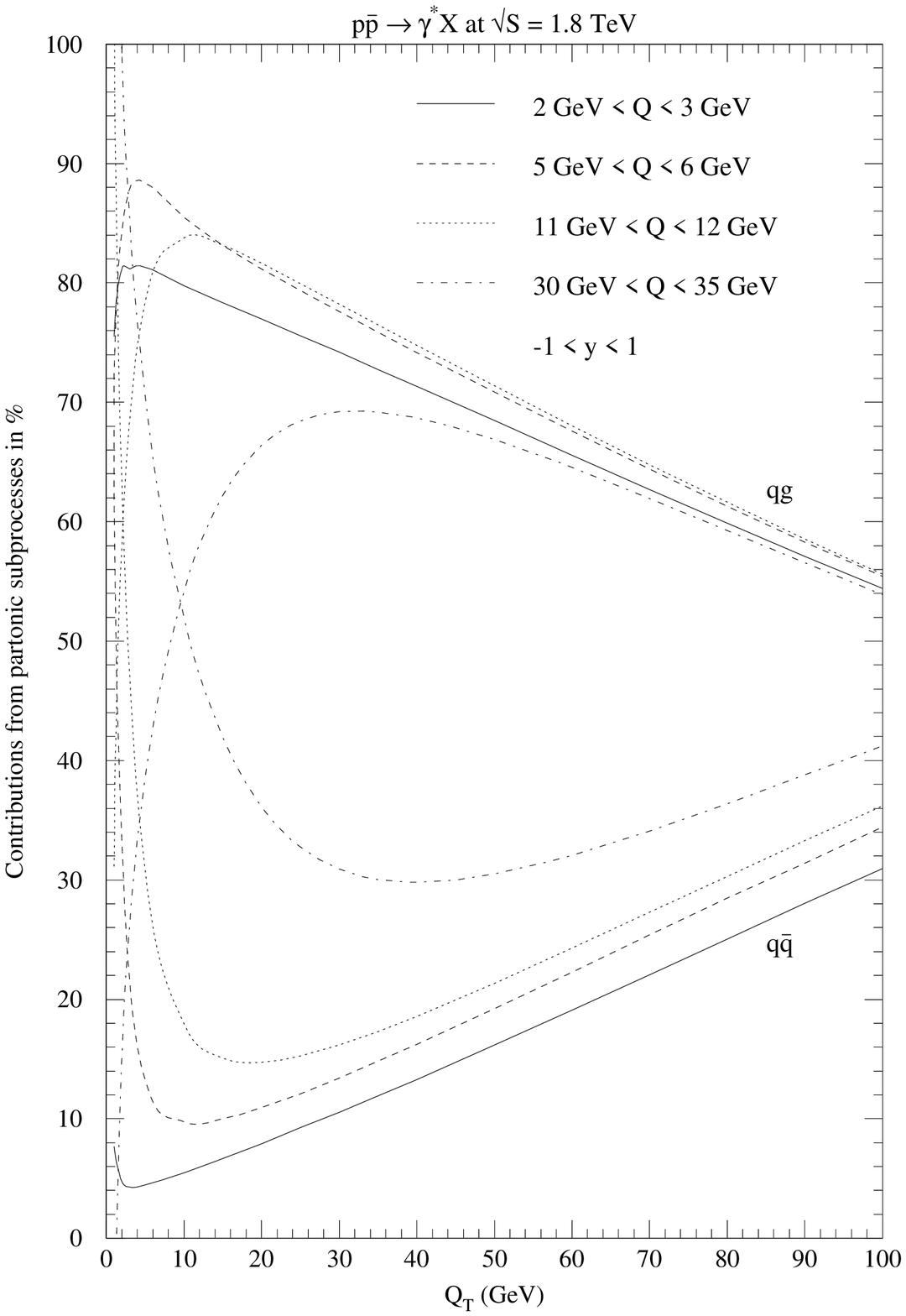}
\vspace*{-0.7cm}
\caption{Contributions from the partonic subprocesses $qg$ and $q \bar{q}$ 
to the invariant 
inclusive cross section $Ed^3\sigma/dp^3$ as a function of $Q_T$ for 
$p {\bar p}\rightarrow \gamma^* X$ at $\protect\sqrt{S}$ =
1.8 TeV.} 
\label{fig:1}
\end{figure}
$\sqrt S =$ 1.8 TeV and several values of the mass of the lepton-pair, 
we present in Fig.~1 the $q {\bar q}$ and $q g$ perturbative contributions 
to the invariant inclusive cross section $Ed^3\sigma/d p^3$ as a function of 
$Q_T$. For small $Q_T$, the $q {\bar q}$ contribution exceeds that of $q g$ 
channel. However, as $Q_T$ grows, the $q g$ contribution becomes 
increasingly important and accounts for 70 to {80 \%} of the rate once 
$Q_T \simeq Q$. (The $q {\bar q}$ contribution begins to be felt a second time 
at very large $Q_T$ owing to the valence nature of the ${\bar q}$ density in 
the ${\bar p}$.)  Subprocesses other than those initiated by the 
$q {\bar q}$ and $q g$ initial channels contribute negligibly.  

Prompt photons have been observed in Fermilab Tevatron collider 
experiments with values of 
$p_T$ extending to 100 GeV and beyond.  Lepton-pair cross sections are 
smaller owing to the factor $\alpha_{em}/(3 \pi Q^2)$ associated with the 
decay of the virtual photon to $\mu^+ \mu^- $.  It should be possible to 
examine massive lepton-pair cross sections in the 
same data sample out to $Q_T$ of 30~GeV or more.  
\begin{figure}
\epsfxsize200pt
\figurebox{}{}{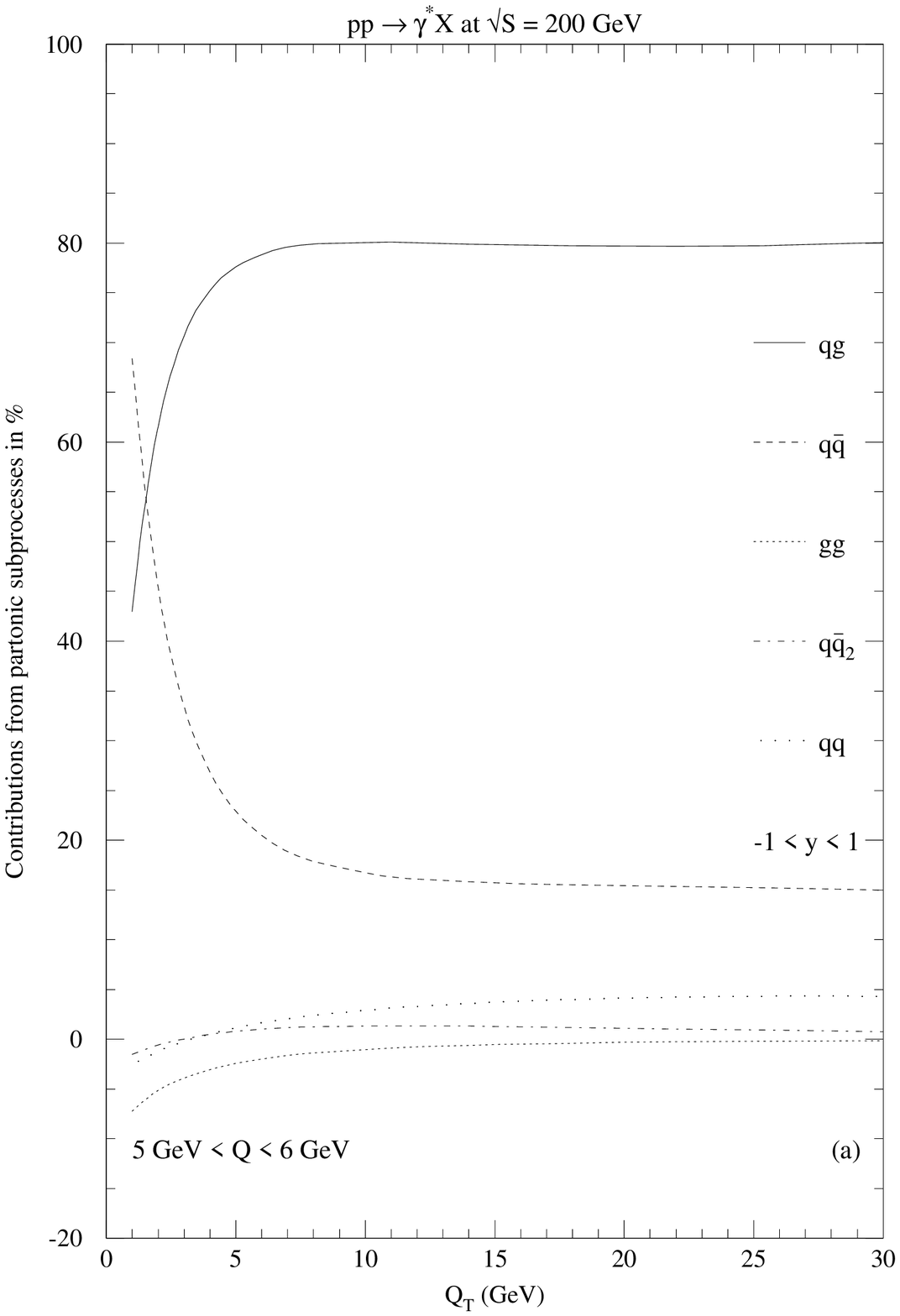}
\vspace*{-0.7cm}
\caption{Contributions from the partonic subprocesses $qg$ and $q \bar{q}$ 
to the invariant inclusive cross section $Ed^3\sigma/dp^3$ as a function of 
$Q_T$ for $p p \rightarrow \gamma^* X$ at $\protect\sqrt{S}$ = 200 GeV.} 
\label{fig:2}
\end{figure}
The statistical limitation to $Q_T$ of about 30 GeV means that the reach 
in $x_{gluon}$, the fractional light-cone 
momentum carried by the incident gluon, is limited presently to 
$2Q_T/\sqrt S \sim 0.033$, a factor of three less than that 
potentially accessible with prompt photons.    
It is valuable nevertheless to investigate the gluon density in the region 
$x_{gluon} \sim 0.033$, and less, with a process that has reduced 
experimental and theoretical systematic uncertainties from those of the 
prompt photon case.  

In Fig.~10 of the first paper\cite{BGKDY}, we show a comparison with data 
of our calculated invariant inclusive cross section $Ed^3\sigma/d p^3$ as a 
function of $Q_T$ for $p + {\bar p} \rightarrow \gamma^* +X$, with 
$\gamma^* \rightarrow \mu^+ \mu^-$, at $\sqrt S =$ 630 GeV, with 
$2m_{\mu} < Q <$ 2.5 GeV.  The theoretical expectation is in good agreement 
with the data published by the CERN UA1 collaboration\cite{expt2}.  Dominance 
of the $qg$ component is evident over a large interval in $Q_T$. It would 
be valuable to make a similar comparison with Tevatron data.  

Results similar to those above are shown in Fig.~2 for 
$p p \rightarrow \gamma^* + X$ at $\sqrt S =$ 200 GeV appropriate 
for the RHIC collider at Brookhaven.  The fraction of the cross 
section attributable to $qg$ initiated subprocesses 
again increases with $Q_T$, growing to {80 \%} for $Q_T \simeq Q$.  
Predictions of spin-averaged and spin-dependent 
cross sections for the energies of the RHIC collider may be found in 
the second paper\cite{BGKDYX}.  
Adopting the nominal value $Ed^3\sigma/dp^3 = 10^{-3} \rm{pb/GeV}^2$, we 
establish that the massive lepton-pair cross section may be measured 
to $Q_T$ =14 and 18.5 GeV in $p p \rightarrow \gamma^* + X$ at 
$\sqrt S =$200 and 500 GeV, respectively, when 2 $< Q <$ 3 GeV, and 
to $Q_T =$11.5 and 15 GeV when 5 $< Q <$ 6 GeV.  In terms of reach 
in $x_{gluon}$, these values of 
$Q_T$ may be converted to $x_{gluon} \simeq x_T = 2 Q_T/\sqrt S =$ 
0.14 and 0.075 at $\sqrt S =$ 200 and 500 GeV when 2 $< Q <$ 3 GeV, and 
to $x_{gluon} \simeq$ 0.115 and 0.06  when 5 $< Q <$ 6 GeV.  The 
smaller cross section in the case of massive lepton-pair 
production means that the reach in $x_{gluon}$ is restricted to a factor of 
about two to three less, depending on $\sqrt S$ and $Q$, than that potentially 
accessible with prompt photons in the same sample of data.  

In the first paper\cite{BGKDY}, we compare our spin-averaged cross 
sections with fixed-target data on massive lepton-pair production at 
large values of $Q_T$, and we establish that fixed-order 
perturbative calculations, without resummation, should be reliable for 
$Q_T > Q/2$.  
 
Although the $qg$ Compton subprocess is dominant, one might question whether
uncertainties associated with the quark density compromise the possibility to
determine the gluon density.  In this context, it is useful to
recall\cite{elbjwq} that when the Compton subprocess is dominant, the 
spin-averaged cross sections for prompt photon production and for 
lepton-pair production may be rewritten in a form in which the quark 
densities do not appear explicitly, but, instead, directly 
in terms of the proton structure function $F_2(x,\mu_f^2)$ {\it measured} 
in spin-averaged deep-inelastic lepton-proton scattering.  An analogous 
statement applies in the spin-dependent case where the lepton pair cross 
section may be expressed in terms of the $g_1(x,\mu_f^2)$ structure function 
measured in spin-dependent deep-inelastic lepton-proton scattering.

\section*{Acknowledgments}
Work at Argonne National Laboratory is supported by the U.S. Department of 
Energy, Division of High Energy Physics, under Contract W-31-109-ENG-38. 
M.K. is supported by DFG through grant KL 1266/1-1.

\end{document}